\documentclass{iopart}
\pdfoutput=1
\usepackage{graphicx,latexsym,iopams,hyperref}
\newcommand{\ket}[1]{\left| #1 \right\rangle}
\newcommand{\bra}[1]{\left\langle #1 \right|}

\newcommand{\equref}[1]{(\ref{#1})}%IOP does not use Eq. ()

\begin{document}

\title{Fractional quantum Hall effect in a U(1)$\times$SU(2) gauge field}
\author{Rebecca N. Palmer and Jiannis K. Pachos}
\address{School of Physics and Astronomy, University of Leeds, Leeds, LS2 9JT, United Kingdom} 
\ead{R.Palmer@leeds.ac.uk}
\date{\today}

\begin{abstract}
We consider the bosonic fractional quantum Hall effect in the presence of a non-Abelian gauge field in addition to the usual Abelian magnetic field.  The non-Abelian field breaks the twofold internal state degeneracy, but preserves the Landau level degeneracy.  Using exact diagonalization, we find that for moderate non-Abelian field strengths the system's behaviour resembles a single internal state quantum Hall system, while for stronger fields there is a phase transition to either two internal state behaviour or the complete absence of fractional quantum Hall plateaus.  Usually the energy gap is reduced by the presence of a non-Abelian field, but some non-Abelian fields appear to slightly increase the gap of the $\nu=1$ and $\nu=3/2$ Read-Rezayi states.
\end{abstract}

\submitto{\NJP}
\maketitle

\section{Introduction} 

The fractional quantum Hall (FQH) effect \cite{fqh-frac-gauge} occurs in two-dimensional interacting systems in a strong perpendicular Abelian gauge field at sufficiently low temperatures. It manifests as incompressible strongly correlated ground states, at several simple fractional values of the filling factor (ratio of particles to flux quanta), with anyonic excitations.  The FQH effect is usually realized with electrons in a magnetic field, but theoretically exists for both bosons and fermions, and both long- and short-ranged interactions \cite{rotating-gas-review,rotating-gas-review2}.  Several methods have been proposed for implementing it with ultracold atoms \cite{artificial-gauge-review,fqh-laserhop,fqh-laserhop2,fqh-bec-immersion,fqh-quadrupole,fqh-stark,fqh-rotation,fqh-oam}, and some few-atom experiments have been done \cite{fqh-few-atom}.

One motivation for studying the FQH effect is that it can support anyons \cite{topo-qc-review,na-anyon-review}. These are particles whose exchange statistics is not the $\pm 1$ of bosons or fermions, but some other phase $\rme^{\rmi\theta}$ for Abelian anyons, or a unitary matrix for non-Abelian anyons.  This statistical effect depends only on the topology of the anyons' paths and not on the details of their motion.  In the case of non-Abelian anyons, the states on which the statistical matrix acts are a global property of the pair, indistinguishable if each anyon is measured individually, but can be measured by bringing them together.  Hence, a quantum computer using well separated non-Abelian anyons as qubits and statistical matrices as gates would be immune to error from either local decoherence or imperfect control of the anyon motion \cite{topo-qc-orig}, as long as there were no uncontrolled stray anyons in the system.  For some types of anyons, including those expected in some FQH states \cite{rr-qc-braid,topo-qc-review}, such a quantum computer would be universal.

However, errors will occur if uncontrolled stray anyons, such as thermally created anyons \cite{thermal-anyon1,thermal-anyon2}, move around the computational anyons.  If the temperature is too high compared to the anyon pair creation gap, these stray anyons will make the system unusable for computation, and eventually destroy the FQH effect completely.  In standard FQH systems, the states supporting non-Abelian anyons have small enough energy gaps that they are difficult to observe \cite{fqh-2ll-expt,mr-interference-expt} and would be even more difficult to use for computation.  This provides a motivation to look for modified FQH systems with more robust non-Abelian anyons.

The modification we consider here is to add a uniform SU(2) non-Abelian component to the magnetic field, coupled to a two-dimensional internal state space of the particles. Such a system can be realized using ultracold atoms in an optical lattice \cite{artificial-gauge-review,u2f-const-iqh,u2f-graphene-iqh,u2f-rashba-zeeman-iqh,u2f-laserhop,u2f-valley,u2f-u3f-darkstate}, or using electrons in a material with spin-orbit splitting \cite{spin-orbit-equal,spin-orbit-levels}.  We study this system using lowest Landau level exact diagonalization, for bosonic particles and a two-parameter range of SU(2) field strengths.  When both parameters are of the same order as the Abelian field strength, the system behaves similarly to a \emph{one} internal state FQH system, i.e. incompressible states at filling factors $\nu=\frac{1}{2},1,\frac{3}{2},\ldots$ with large overlaps with the Read-Rezayi states \cite{rr-torus-numeric}, despite actually having two internal states.  For most parameter values, these have smaller energy gaps than in a true one internal state system, but in the cases $\nu=1,\frac{3}{2}$, it appears that some choices of non-Abelian field can slightly increase the gap.  When both parameters are large, we find no incompressible states at all, while when only one is large, we find similar behaviour to when both are zero, i.e. a two internal state FQH system.

In section \ref{model} we introduce our model, starting from a lattice Hamiltonian, then taking the continuum limit. In subsection \ref{sec1part} we derive the single particle Landau levels in the presence of the SU(2) gauge field. Section \ref{numerics} gives many-particle exact diagonalization results for the density profile, the energy gap, and the overlap with some trial states.  In particular, we consider the $\nu=1/2$ Laughlin state and the $\nu=1$ Moore-Read state.  We conclude in section \ref{conc}.

\section{The Model}\label{model}
We consider bosons in two dimensions, in either continuous space or a lattice that can be well approximated by continuous space.  We apply a uniform classical U(2)=U(1)$\times$SU(2) gauge field, minimally coupled to two internal states of the particles.  The U(1) part is the Abelian field of standard FQH, while the SU(2) non-Abelian part is our new ingredient.  The FQH effect also requires an interaction between the particles, which we take to be a contact interaction.  A possible physical implementation of this system is described in \cite{u2f-const-iqh,artificial-gauge-review}, using ultracold bosonic atoms in an optical lattice with laser induced hopping.

\subsection{Lattice Hamiltonian}
We consider bosons with two internal states on a square lattice, described by the Bose-Hubbard Hamiltonian with a classical gauge field \cite{artificial-gauge-review},
\begin{eqnarray}
\fl H=\sum\limits_{x,y}-J\left[a^\dagger_{x,y}\exp\left(\rmi\int\limits_{x-1}^x\bi{A}(x^\prime,y)dx^\prime\right)a_{x-1,y}\right.\nonumber\\
\left.+a^\dagger_{x,y}\exp\left(\rmi\int\limits_{y-1}^y\bi{A}(x,y^\prime)dy^\prime\right)a_{x,y-1}+\mathrm{H.c.}\right]\nonumber\\
+Ua^\dagger_{x,y}a^\dagger_{x,y}a_{x,y}a_{x,y}.\label{Hgauge}
\end{eqnarray}
Here $\bi{A}(x,y)$ is the vector potential of the gauge field, $a^\dagger_{x,y}$ creates a particle in lattice site $(x,y)$ (we use the lattice spacing as the unit of length), $J$ is the hopping rate and $U$ is the interaction strength.  For simplicity, we assume isotropic hopping, and state-independent hopping and interaction.

In the case of a non-Abelian gauge field, the particles have multiple internal (e.g. hyperfine) states, and $\bi{A}$ is a matrix in internal state space as well as a vector field in real space, i.e. hopping can change the particle's internal state.  We consider the case of a uniform U(2)=U(1)$\times$ SU(2) gauge field \cite{u2f-const-iqh},
\begin{eqnarray}
\bi{A}(x,y)&=(\alpha\sigma_y,\beta\sigma_x+2\pi\Phi x)\nonumber\\
&=\left(\left(\begin{array}{cc}0&-\rmi\alpha\\\rmi\alpha&0\end{array}\right),\left(\begin{array}{cc}2\pi\Phi x&\beta\\\beta&2\pi\Phi x\end{array}\right)\right),
\label{gauge}
\end{eqnarray}
where the $\sigma_i$ are the Pauli matrices.  The Abelian part of this field is parameterized by $\Phi$, and the non-Abelian part by $\alpha,\beta$.  As we see below, the Abelian part is necessary to have an FQH effect; a pure SU(2) field \cite{su2-bec} does not produce Landau levels.

For this choice of field, $A_x$ is independent of $x$ and $A_y$ of $y$, so the integrals in \equref{Hgauge} simplify to $\rme^{\rmi A_x}$, $\rme^{\rmi A_y}$, giving
\begin{eqnarray}
\fl H=\sum\limits_{x,y}-J\left[a^\dagger_{x,y}(\cos\alpha+\rmi\sigma_y\sin\alpha)a_{x-1,y}\right.\nonumber\\
\left.+\rme^{2\pi \rmi\Phi x}a^\dagger_{x,y}(\cos\beta+\rmi\sigma_x\sin\beta)a_{x,y-1}+\mathrm{H.c.}\right]\nonumber\\
+Ua^\dagger_{x,y}a^\dagger_{x,y}a_{x,y}a_{x,y}.
\label{Hlattice}
\end{eqnarray}

\subsection{Continuum approximation}

When the lattice spacing is small compared to the magnetic length (defined below) and the typical interparticle spacing, we can approximate the site operators $a_{x,y}$ by a continuous field $\psi(x,y)$. Taylor expanding $\psi(x-1,y)$ and similar terms in \equref{Hlattice} about $(x,y)$, and discarding the (dynamically irrelevant) constant $-4J\psi^\dagger\psi$, gives the approximate Hamiltonian
\begin{eqnarray}
\fl H\approx H_c\equiv\int dxdy J\psi^\dagger\left[\left(-\rmi\frac{\partial}{\partial x}+\alpha\sigma_y\right)^2\right.\nonumber\\
+\left.\left(-\rmi\frac{\partial}{\partial y}+\beta\sigma_x+2\pi\Phi x\right)^2\right]\psi +\frac{1}{2}U\psi^\dagger\psi^\dagger\psi\psi,\label{Hcont}
\end{eqnarray}
where all occurrences of $\psi$ are taken at position $(x,y)$.
This Hamiltonian has a magnetic length $l=1/\sqrt{2\pi\Phi}$, which sets the length scale of the quantum Hall physics. Hence, the above continuum approximation is valid in the limit $l\gg 1$. For convenience we also define the dimensionless parameters $a\equiv\alpha/\sqrt{\Phi}$ and $b\equiv\beta/\sqrt{\Phi}$.  Within this continuum limit, varying $\Phi$ at fixed $a,b$ changes only the overall length and energy scales.  We can assume without loss of generality that $a,b,\Phi\geq 0$, as a spin space rotation by $\sigma_x$ ($\sigma_y$) changes the sign of $a$ ($b$), while a real space reflection in the $y$ axis changes the signs of $a$ and $\Phi$.

Expanding out the brackets and discarding the dynamically irrelevant constants $\sigma_x^2=\sigma_y^2=I$, we find that the non-interacting part of \equref{Hcont} is equal to the Hamiltonian of 2D electrons with spin-orbit coupling \cite{spin-orbit-levels}, with the coordinate axes rotated by $45^\circ$ (in both real and spin space).  The SU(2) terms correspond to the spin-orbit interaction, which is pure Rashba if $\alpha=-\beta$, pure Dresselhaus if $\alpha=\beta$, and a combination of the two for all other ratios.  An optical lattice loaded with fermionic atoms can hence be used as a quantum simulator of spin-orbit coupled electrons \cite{artificial-gauge-review}.  However, we consider only the bosonic case.
\subsection{Single particle degeneracy: Landau levels}\label{sec1part}

In this subsection we consider the behaviour of a single particle subject to the Hamiltonian \equref{Hcont}. Since $H_c$ is independent of $y$, its single particle eigenstates are plane waves in the $y$ direction, $\psi\propto\rme^{\rmi ky}$.  Substituting this into \equref{Hcont}, we find that the eigenstates have the form $\psi=\rme^{\rmi ky}\Psi_n(x+k/2\pi\Phi)$, where $\Psi_n$ and the energy $E_n$ satisfy the equation
\begin{eqnarray}
\fl E_n\Psi_n(X)=J[(-\rmi\partial/\partial X+\alpha\sigma_y)^2\nonumber\\
+(\beta\sigma_x+2\pi\Phi X)^2]\Psi_n(X).\label{lll-wavefn}
\end{eqnarray}
Like the Abelian case, the energy is hence independent of $k$, so each single particle energy level $n$ (called a Landau level) has a macroscopic degeneracy of $\Phi$ states per unit area (one per Abelian flux quantum).  In the many-particle case, this degeneracy is broken by the contact interaction, so a weak interaction can produce strongly correlated states.  As in standard FQH, we define the filling factor $\nu$ as the number of particles divided by the number of states in a Landau level, $\nu\equiv\rho/\Phi$ where $\rho$ is the 2D density.

In the Abelian case ($\alpha=\beta=0$), $E_n=2,2,6,6,10,10,\ldots\times\pi J\Phi$, the additional 2-fold degeneracy coming from interchange of internal states.  In contrast, in the non-Abelian case, the Landau level energies $E_n$ are in general all distinct and follow no simple pattern.  The $E_n$ can be found analytically \cite{spin-orbit-levels}, but this result is complicated, and gives the corresponding wavefunctions $\Psi_n$ only as infinite series; we instead solve \equref{lll-wavefn} numerically.  Full analytic solutions exist in the special cases $\alpha=0$ or $\beta=0$ (2-fold degenerate \cite{spin-orbit-equal}), or $\alpha=\pm\beta$ (non-degenerate \cite{spin-orbit-Rashba,u2f-naqh-symm}).
\begin{figure}
\includegraphics[width=8.6cm]{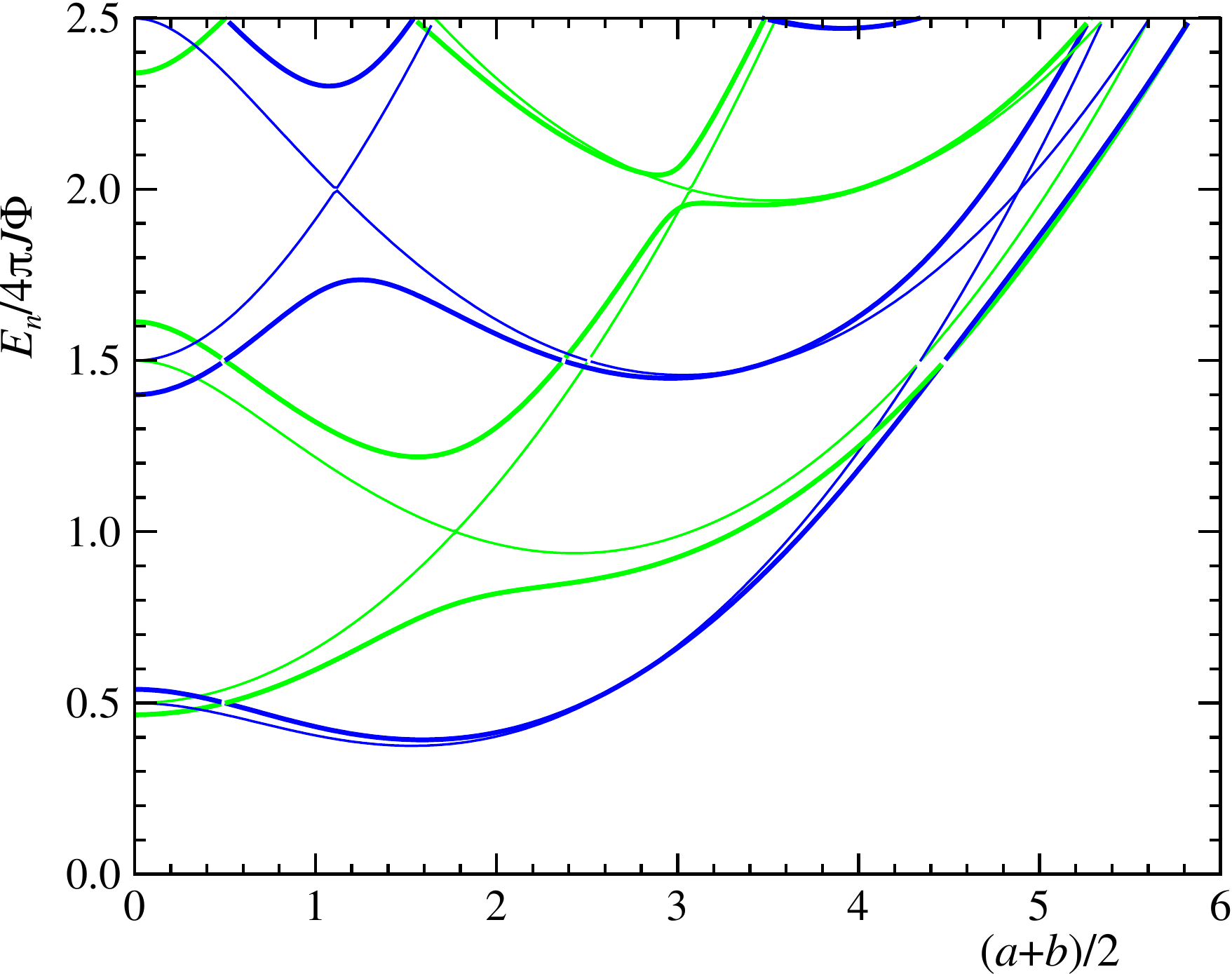}
\caption{Lowest few Landau level energies $E_n$, and symmetries $+-$ (blue/dark) or $-+$ (green/light), for $a=b$ (thin lines) and $a=b\pm 1$ (thick lines).} \label{llevels}
\end{figure}

\begin{figure}
\includegraphics[width=8.6cm]{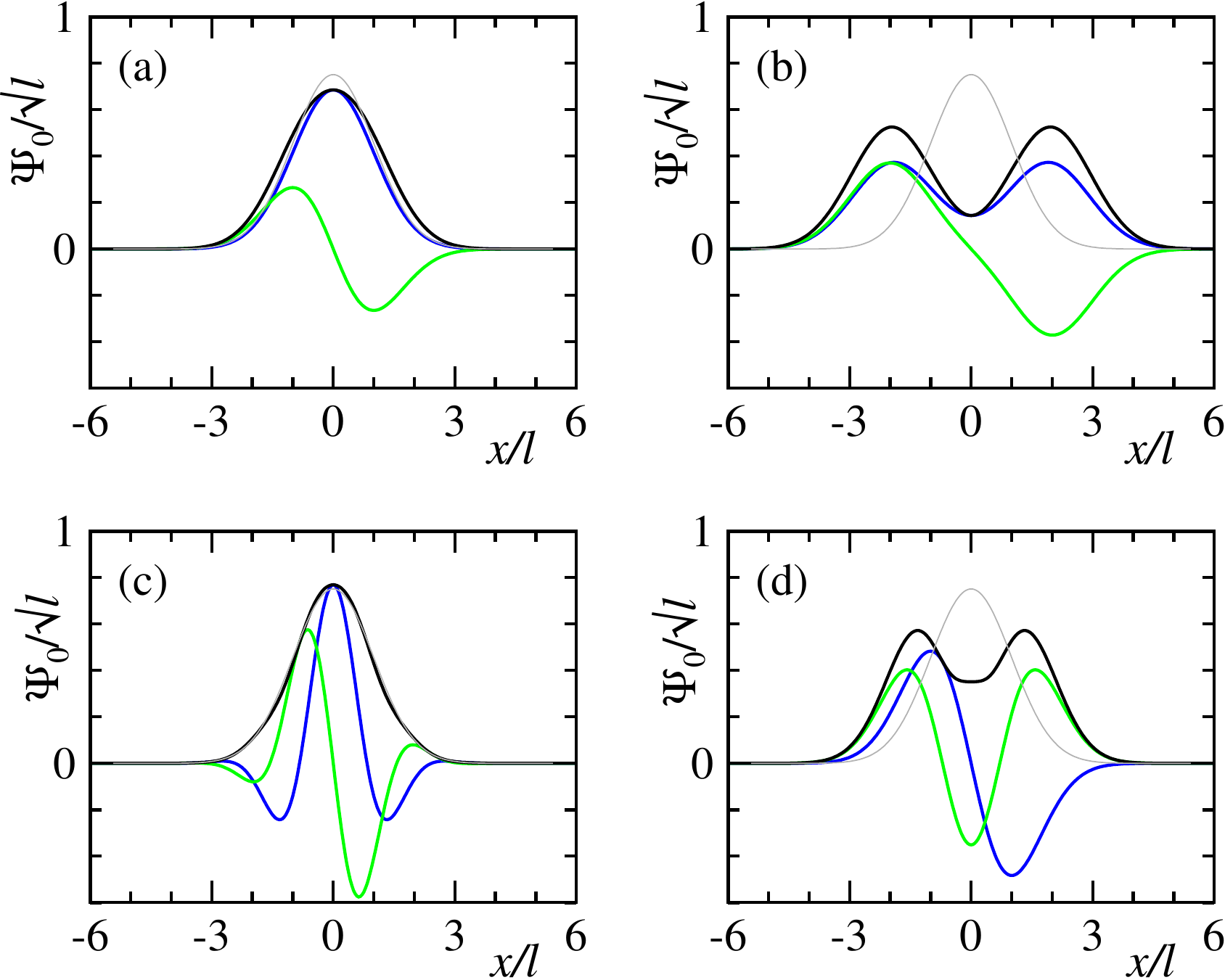}
\caption{Single particle lowest Landau level wavefunctions $\Psi_0(x)$, for non-Abelian field strengths $(a,b)=$ (a) (1,1), (b) (1,5), (c) (5,1), (d) (5,5).  Blue and green are the two internal states $\Psi_0(x,\uparrow)$, $\Psi_0(x,\downarrow)$, black their rms total amplitude $\sqrt{|\Psi_0(x,\uparrow)|^2+|\Psi_0(x,\downarrow)|^2}$, thin grey the Abelian-field ground state for comparison.} \label{1part-wavefn}
\end{figure}

\begin{figure}
\includegraphics[width=8.6cm]{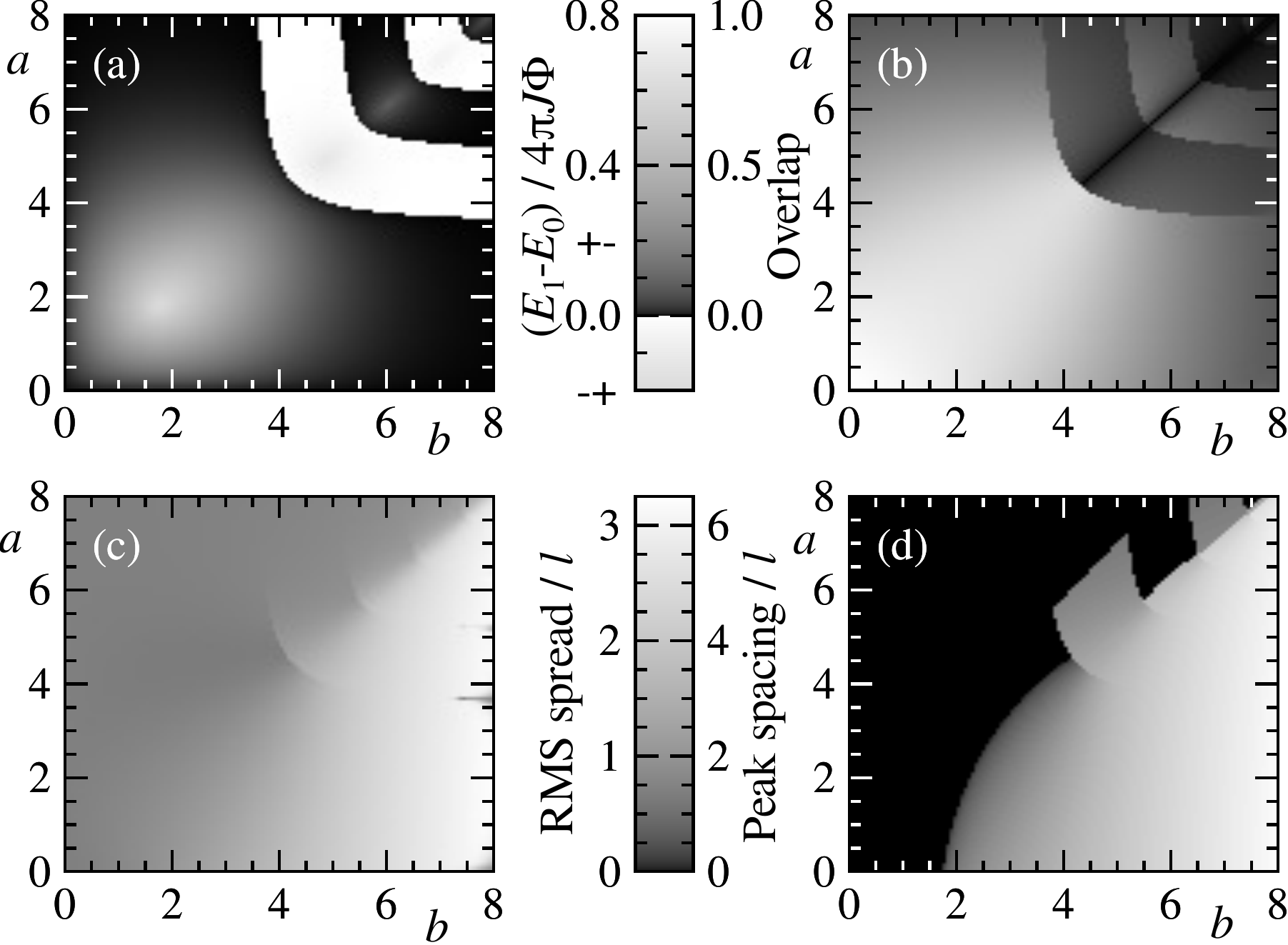}
\caption{Single particle state properties: (a) Energy difference $(E_1-E_0)/4\pi J\Phi$ between the lowest two Landau levels, and symmetry ($+-$ or $-+$) of the lowest level.  (b) Overlap $\langle\Psi_{0,\mathrm{Abelian}}|\Psi_0\rangle$ of the lowest Landau level wavefunction with the Abelian lowest Landau level.  (c) RMS spread (standard deviation) of the lowest Landau level wavefunction, $(\bra{\Psi_0}x^2\ket{\Psi_0})^{1/2}/l$.  The small dark features on the right hand edge are numerical artefacts, caused by the gap between the two lowest Landau levels being too small to resolve.  (d) Distance between the two highest maxima of $\Psi_0$ (equal height by symmetry, see figure \ref{1part-wavefn}b,d), or zero if there is a single central maximum.} \label{1part}
\end{figure}

Figure \ref{llevels} plots the lowest few Landau level energies $E_n$ along two diagonal lines in $(a,b)$ space, while figure \ref{1part}a plots the difference between the lowest two energies against $a,b$.  Figure \ref{1part-wavefn} plots the lowest Landau level wavefunction $\Psi_0$ at four $(a,b)$ settings.  Because \equref{lll-wavefn} has the $Z_2$ symmetry $\Psi_n(x)\rightarrow\sigma_z \Psi_n(-x)$, in each Landau level either the first internal state component is symmetric about $x=-k/2\pi\Phi$ and the second antisymmetric ($+-$ symmetry, figure \ref{1part-wavefn}a,b,c), or vice versa ($-+$, figure \ref{1part-wavefn}d).  Levels of the same symmetry avoided cross, touching only in conical intersections on the $a=\pm b$ diagonals, but levels of opposite symmetry cross freely, on lines in $(a,b)$ space.  For small positive $a,b$, the lowest Landau level is $+-$ and the second $-+$, but at larger $a,b$ these cross infinitely many times \cite{u2f-naqh-symm}, causing a discontinuous change in the wavefunction $\Psi_0$ each time.

The exactly solvable cases $a=0$ and $b=0$ can be used as perturbative approximations for the cases when $a$ or $b$ is small but not zero \cite{spin-orbit-equal}.  When $a=0$, the two degenerate ground states at a given $k$ are Gaussians with shifted centres, $\psi(x,y)=\exp\{\rmi k y-[x+(k\pm\beta)/2\pi\Phi]^2/2l^2\}$, in the $\sigma_x=\pm 1$ internal states (equal superpositions with relative phase $\pm 1$).  When $a\ll b$, the ground state is close to the symmetric combination of these (figure \ref{1part-wavefn}b), and hence has two peaks approximately $\beta/\pi\Phi=bl/\sqrt{\pi}$ apart (figure \ref{1part}d).  When $b=0$, the ground states have a Gaussian amplitude but an extra phase, $\psi(x,y)=\exp[\rmi (ky\pm\alpha x)-(x+k/2\pi\Phi)^2/2l^2]$ with internal states $\sigma_y=\pm 1$.  The symmetric combination, and approximate $a\gg b$ ground state (figure \ref{1part-wavefn}c), has a Gaussian total amplitude but a position dependent internal state.

However, this apparent asymmetry between $a$ and $b$ arises entirely from our choice to have states extended along $y$ and localized in $x$, and to use a gauge that simplifies such states; we could instead have made the Abelian part of $\bi{A}$ proportional to $y$ (by a gauge transformation) and had states $\psi\propto\rme^{\rmi kx}$, which would interchange the roles of $a$ and $b$.

\section{Numerical results}\label{numerics}

We now consider the multi-particle case.  We assume weak interaction, by which we mean that the interaction energy per particle ($\sim U\Phi$ at $\nu\sim 1$) is small compared to the Landau level spacing ($\sim J\Phi$), i.e. $U\ll J$.  We can hence assume that all the particles are in the lowest Landau level, and use a Fock basis over the single particle eigenstates $\ket{k}$.  Since all momenta $k$ within a Landau level are degenerate, the only non-trivial term is the interaction term, with the matrix elements
\begin{eqnarray}
\fl\bra{k_1^\mathrm{out},k_2^\mathrm{out}}H_\mathrm{int}\ket{k_1^\mathrm{in},k_2^\mathrm{in}}=U\delta(k_1^\mathrm{in}+k_2^\mathrm{in}-k_1^\mathrm{out}-k_2^\mathrm{out})\nonumber\\
\times\int dx \Psi_0^*(x+k_1^\mathrm{out}/2\pi\Phi)\Psi_0^*(x+k_2^\mathrm{out}/2\pi\Phi)\nonumber\\
\times\Psi_0(x+k_1^\mathrm{in}/2\pi\Phi)\Psi_0(x+k_2^\mathrm{in}/2\pi\Phi).
\end{eqnarray}
We adopt periodic boundary conditions (a torus geometry) \cite{qh-torus-llstates,qh-torus-2dsymm}, which makes the number of allowed $k$ finite, and use the full 2D Haldane symmetry \cite{qh-torus-2dsymm}.  The interaction Hamiltonian then becomes a finite matrix, which we diagonalize to find the ground and lowest few excited states.  This method is exact within the lowest Landau level, and in particular can describe the strong correlations typical of FQH states, but is limited to small particle numbers by the exponentially growing basis size.

\subsection{Density profiles}\label{secdensity}
\begin{figure}
\includegraphics[width=8.6cm]{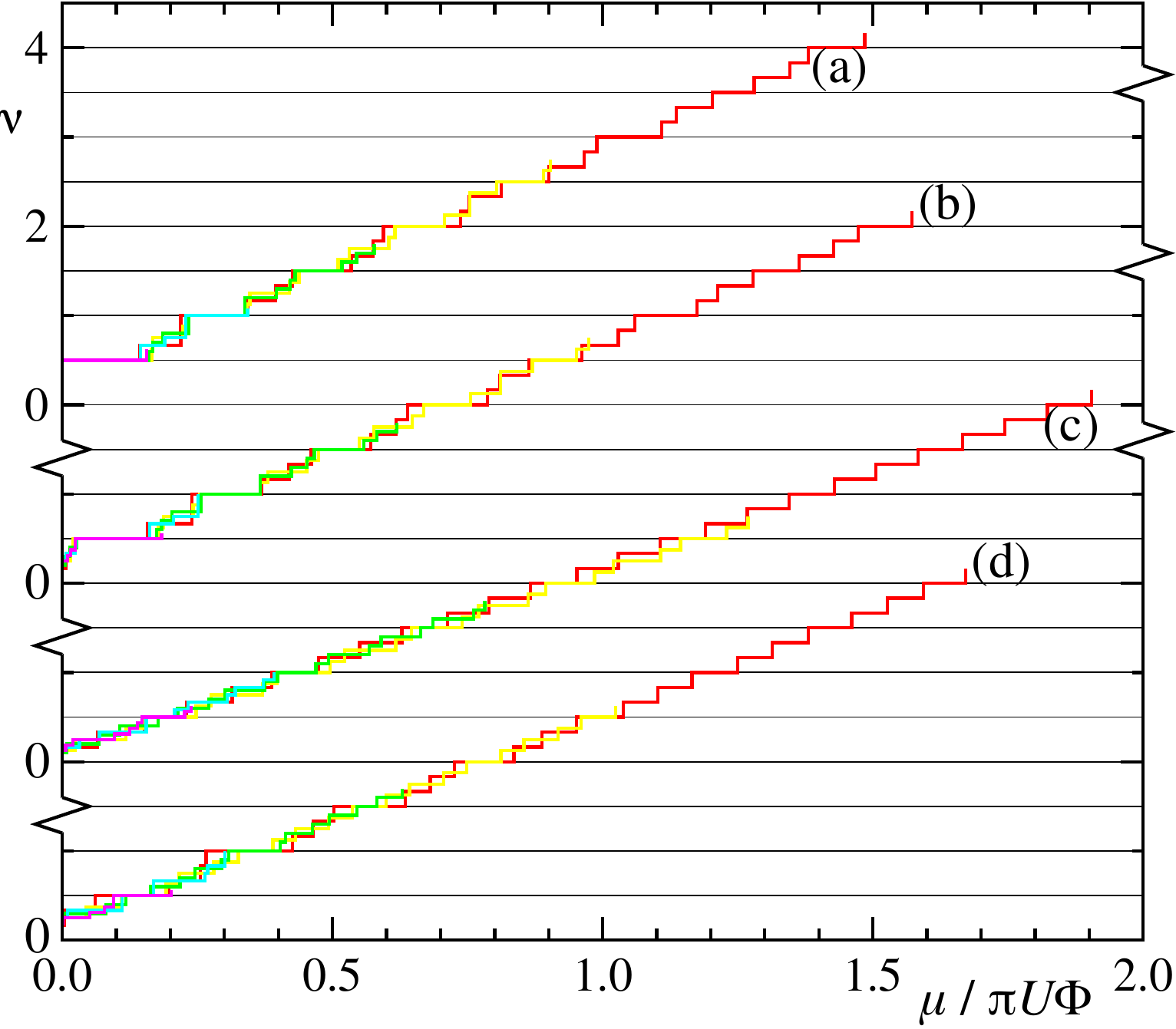}
\caption{Filling factor (zeros offset for clarity) against chemical potential, found by exact diagonalization on a square torus.  (a) Abelian field (single internal state), (b-d) non-Abelian fields $(a,b)=$ (b) (2,2), (c) (5,5), (d) (5,1).  System sizes 6 (red), 8 (yellow), 10 (green), 12 (cyan), 16 (magenta) flux quanta; due to computational limitations, the larger sizes can only be used at small filling factors.  Horizontal lines mark the Read-Rezayi densities $\nu=1/2,1,3/2,\ldots$.} \label{density}
\end{figure}\begin{figure}
\includegraphics[width=8.6cm]{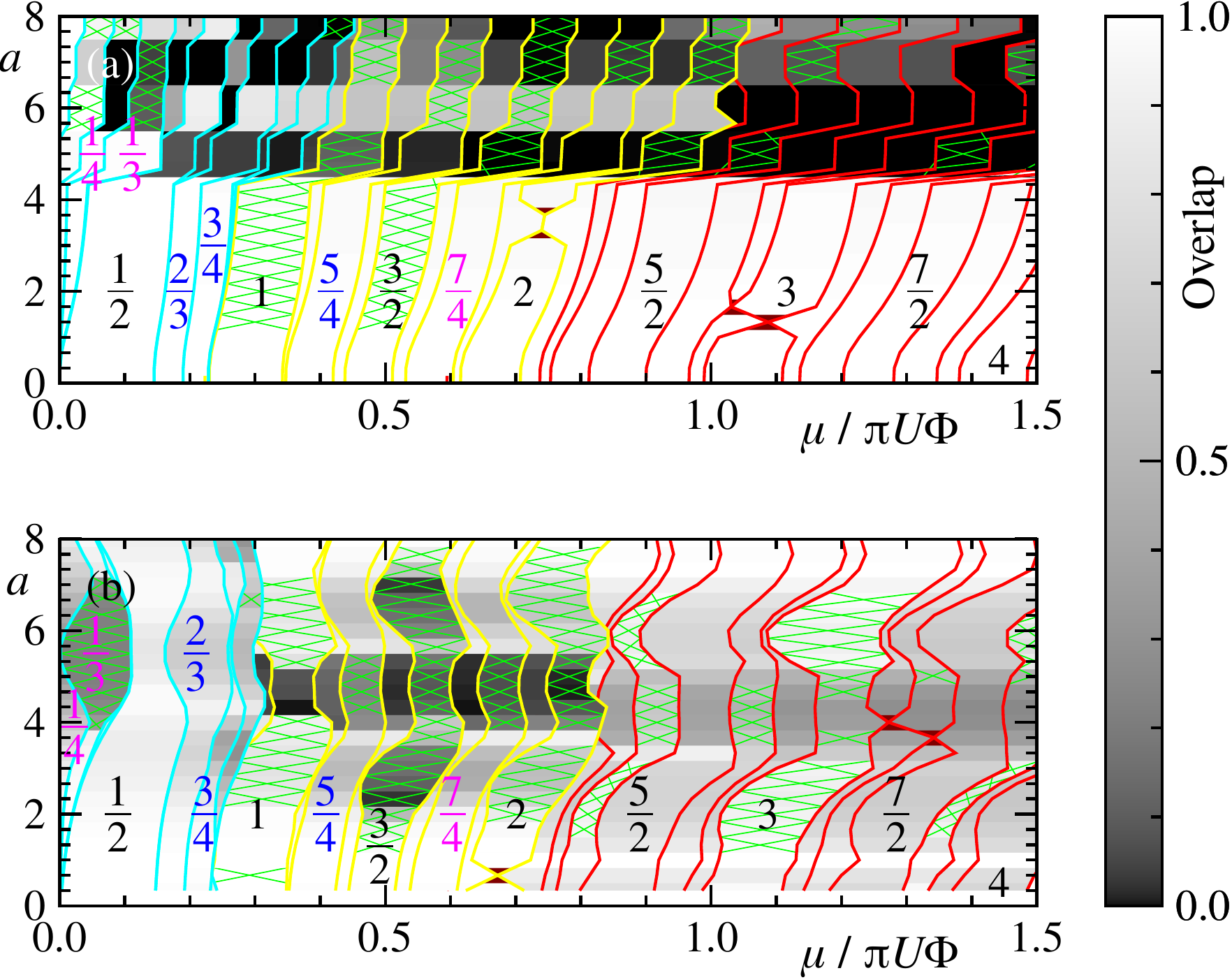}
\caption{Filling factor (lines mark plateau edges, Read-Rezayi series $\nu=p/2$ labelled in black text, composite fermion series $\nu=p/(p\pm 1)$ in blue, others in magenta), and projected overlap with the corresponding Abelian-field state (shading), against chemical potential and $a$.  Overlap is with the lowest Abelian state of the same Haldane momentum $\bi{K}$, projected into the non-Abelian lowest Landau level; green crosses indicate that this is not the same $\bi{K}$ as the Abelian ground state.  (a) $a=b$, (b) $b=1$.  System sizes 12 ($\nu<1$, cyan lines), 8 ($1\leq\nu\leq 2$, yellow), 6 ($\nu>2$, red) flux quanta, on a square torus.  Dark red triangles indicate densities skipped because the exact diagonalization failed to converge.} \label{density2}
\end{figure}

We perform exact numerical diagonalization for different numbers of particles (at a fixed number of Abelian flux quanta, i.e. number of single particle states per Landau level) to obtain the ground state energy as a function of density.  We then convert this to density as a function of chemical potential \cite{fqh-density-steps}.  This approach makes incompressible states visible as plateaus in the density.  Furthermore, in the local density approximation (valid for weak traps) the local chemical potential equals the global chemical potential minus the trap potential \cite{fqh-density-steps}, so these results can be directly compared with experiment if the in-trap density profile can be measured.

At moderate non-Abelian field strengths ($a,b\lesssim 4$), the density profile (figure \ref{density}b, lower halves of figures \ref{density2}a,b) resembles the {\em single} internal state Abelian-field case (figure \ref{density}a and \cite{fqh-density-steps}).  Furthermore, most of the ground states have large ($>90\%$) overlaps with the projections of their Abelian-field equivalents onto the new lowest Landau level.  This makes sense, as though we have two internal states, the lowest Landau level is a single level, not a degenerate pair.  The main plateaus are the Read-Rezayi series $\nu=p/2$ ($p$ integer), clearly present for $\nu=1/2,1,\ldots,3$ and possibly present all the way up to our calculation's limit of $\nu=4$.  (Because the number of particles must be an integer, finite system density profiles are stepped even in compressible phases, so narrow incompressible plateaus cannot be reliably distinguished from regions of low but non-zero compressibility.)  Of the composite fermion series $\nu=p/(p\pm 1)$, we see hints of $\nu=2/3,3/4,4/5,7/6,6/5,5/4$ (but not $5/6$ or $4/3$).  We also see $\nu=7/4$, $9/4$ and $11/4$, but do not have any theory as to what these states are (the previous Abelian-field study \cite{fqh-density-steps} used 6 flux quanta, at which these $\nu$ are not possible).  They may be finite size artefacts, as the $5/4$ plateau narrows greatly on increasing the system size from 8 to 12 states.  Due to computational limitations, we were unable to do this test for the higher densities.  There are two main differences from the Abelian case at these values of $a,b$.  First, all plateaus are shifted towards higher chemical potential.  This would cause a trapped system (with fixed particle number) to slightly expand. Second, the Laughlin state ($\nu=1/2$) no longer has exactly zero interaction energy, and possible new plateaus appear below it at $\nu=1/4,1/3$, though again these are too narrow to be sure they are real.

If the non-Abelian field strength is further increased along the $a=b$ diagonal (figure \ref{density}c, upper half of figure \ref{density2}a), all the plateaus disappear suddenly, in a first order phase transition. As this happens at the same point as the single particle Landau levels cross (figure \ref{llevels}), it is probably caused by this sudden change of single particle states.

If $a$ is increased while $b$ is kept constant (figure \ref{density}d, upper half of figure \ref{density2}b) or vice versa, most of the plateaus disappear gradually, though at some system sizes a few plateaus survive or even grow ($\nu=1/3,2/3,1$ for 12 flux quanta), then at even larger $a$ the normal plateaus reappear.  This process becomes slower and occurs at a larger $a$ as the system size is increased, with the minimum plateau width occuring very close to the point where the single particle peak separation equals half the torus circumference ($a=4.1,4.6,5.5$ for 6,8,12 particles and $b=1$, figure \ref{1part}d).  We hence believe the revival, and possibly some of the initial decrease, to be a finite size artefact, caused by single particle states wrapping right round the torus.  In the infinite size limit, it is not clear whether this will become a second order phase transition to a gapless phase at finite $a$ (as suggested by simple $1/N$ extrapolation, figures \ref{laughlin_varyn}Cb, \ref{read_varyn}Cd), or a decay that tends to zero only in the $a\rightarrow\infty$ limit.

However, as the gap between the lowest two Landau levels is exponentially decreasing in this region ($\sim 5\%$ of its peak value at the point of minimum plateau width, figure \ref{1part}a), the weak interaction assumption must eventually fail, allowing the second Landau level to be populated.  In the $a\gg b$ or $b\gg a$ limit, the two Landau levels become degenerate, and by taking superpositions of them we can recover two sets of single particle states with Gaussian amplitudes.  The two states centred at $x=x_0$ are $\psi(x,y)=\exp(-(x-x_0)^2/2l^2-2\rmi x_0 y/l^2\mp\rmi\alpha x)$ ($a\gg b$, internal states $\sigma_y=\pm 1$) or $\psi(x,y)=\exp(-(x-x_0)^2/2l^2-2\rmi x_0 y/l^2\mp\rmi\beta y)$ ($a\ll b$, internal states $\sigma_x=\pm 1$).  Because the interaction does not change the internal state, the extra phase factors cancel out in the interaction Hamiltonian $\psi^\dagger\psi^\dagger\psi\psi$, so in this limit the Hamiltonian matrix is identical to that at $a=b=0$, i.e. Abelian-field FQH with \emph{two} internal states.  This is known to have a $\nu=2/3$ 221 state \cite{rotating-gas-review}, and is conjectured to have a $\nu=2p/3$ NASS series \cite{nass-def}.

The weak interaction assumption would also fail close to Landau level crossings \cite{u2f-naqh-symm}, but here the two degenerate single particle states are not close to Gaussian.  We hence do not know whether these parameter values would be in a two internal state FQH phase, or a gapless phase similar to the one internal state gapless phase above the first such crossing.

\subsection{$\nu=1/2$: Laughlin state}
\begin{figure}
\includegraphics[width=10cm]{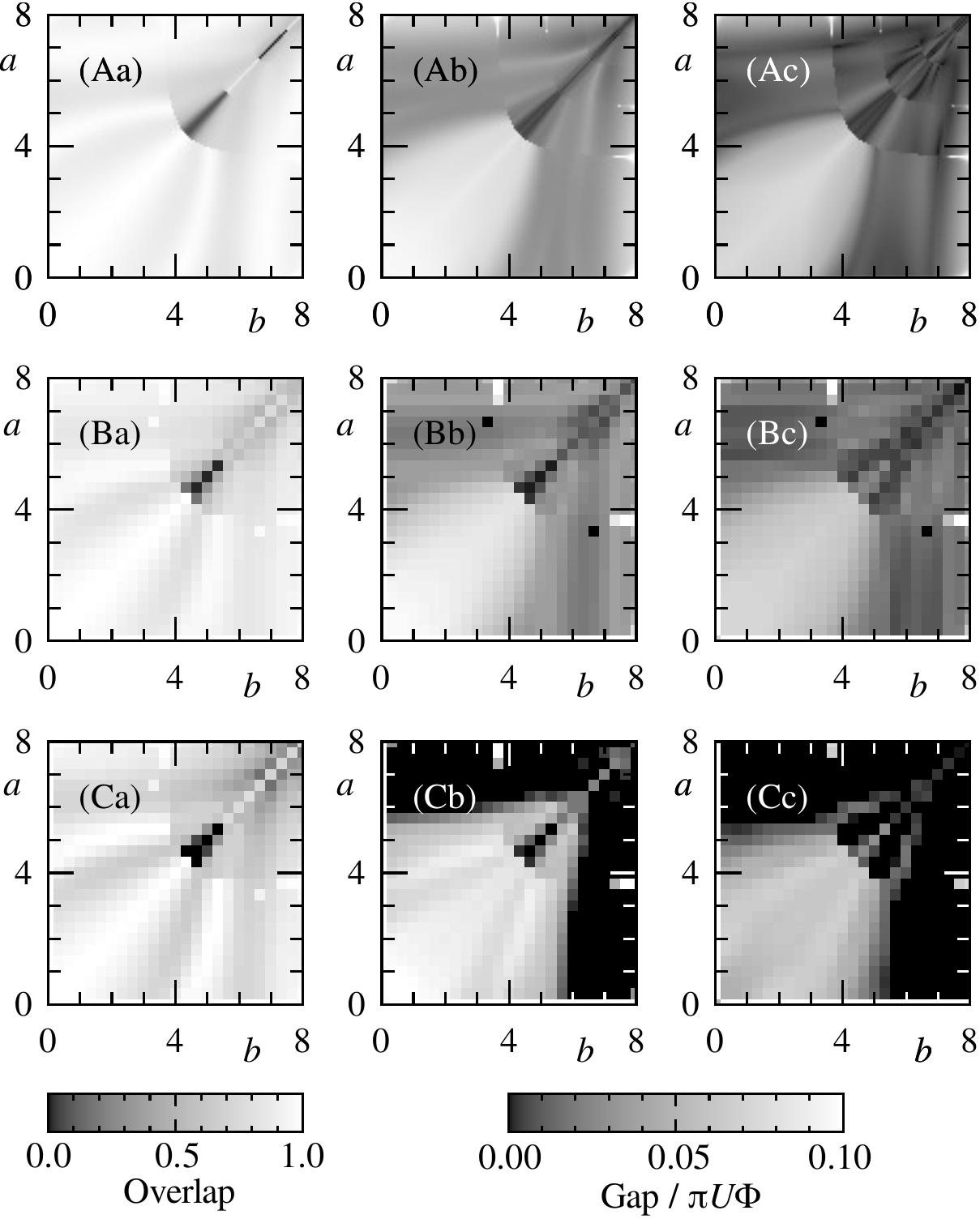}
\caption{$\nu=1/2$, (A) $N=6$ particles, (B) $N=8$ particles, (C) $1/N$ extrapolation to infinitely many particles, based on data for 4-8 particles.  (a) Overlap of the numerically calculated lowest $\bi{K}=(0,0)$ state with the projection of the exact Laughlin state onto the lowest Landau level.  (b) Energy gap between this state and the first excited state (of any $\bi{K}$).  (c) Energy gap between the excitation branch and the continuum (as shown in figure \ref{energies_k}a), smallest gap of $\bi{K}=(N/2,0),(0,N/2),(N/2,N/2)$ ($\bi{K}$ scaled to take integer values $0,\ldots,N-1$).  The small features near $(a,b)=(4,8),(8,4)$ are numerical artefacts.} \label{laughlin_varyn}
\end{figure}
\begin{figure}
\includegraphics[width=8.6cm]{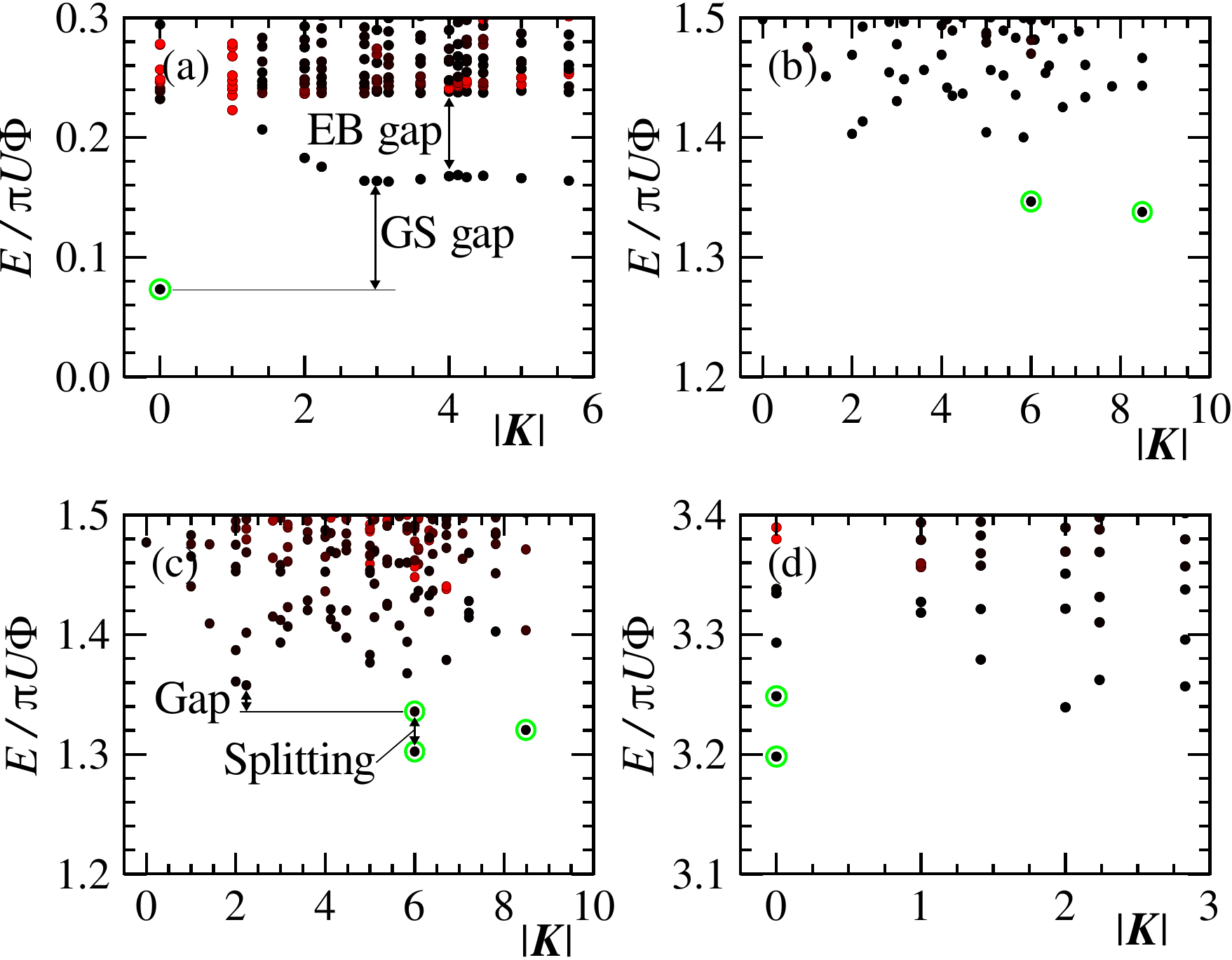}
\caption{Lowest few energies found by exact diagonalization, plotted against $|\bi{K}|$.  (a) $\nu=1/2$ (8 particles, 16 flux quanta), $(a,b)=(2,2)$.  (b,c) $\nu=1$ (12 particles, 12 flux quanta), $(a,b)=$ (b)$(2,2)$, (c)$(2,1)$. (d) $\nu=3/2$ (15 particles, 10 flux quanta), $(a,b)=(2,2)$.  Colour denotes projected overlap with the corresponding Abelian-field state, from none (red) to full (black).  The green circles mark the ground state(s) predicted by Read-Rezayi theory.} \label{energies_k}
\end{figure}

We now consider the $\nu=1/2,1$ states in more detail. The bosonic Laughlin state,
\begin{equation}
\psi(z_1,\ldots,z_N)=\prod\limits_{i,j}(z_i-z_j)^2\rme^{-\sum_i|z_i|^2/4l^2}
\end{equation}
where $z_i=x_i+\rmi y_i$ is the position of particle $i$ as a complex number, has exactly zero interaction energy as it is zero whenever two particles come together.  In Abelian field, it is also a lowest Landau level state, so is the exact ground state at $\nu=1/2$.  It is a non-degenerate state (except for the 2-fold centre of mass degeneracy of all states at half-integer filling \cite{qh-torus-2dsymm}) with zero Haldane momentum, with a gap to a well-defined branch of excitations \cite{rotating-gas-review}.

In non-Abelian field, the Laughlin state is no longer a lowest Landau level state, but it can be projected onto the new lowest Landau level and used as a trial state, which has large overlap with the ground state for nearly all $a,b$ (figure \ref{laughlin_varyn}a).  Its interaction energy is no longer zero after this projection, but for $a,b\lesssim 4$ the energy spectrum remains qualitatively similar (figure \ref{energies_k}a), retaining the gap (figure \ref{laughlin_varyn}b) and well defined excitation branch (figure \ref{laughlin_varyn}c).

At stronger fields, if $a\approx b$ the gap suddenly disappears, while for $a\gg b$ or $a\ll b$ it continuously decreases to near zero then increases again.  As discussed in subsection \ref{secdensity} for general $\nu$, the first is probably a first order phase transition driven by the sudden change of single particle state, while in the second case, it is not clear whether or not the gap actually reaches zero at finite field strength, while the revival is probably a finite size artefact caused by single particle states wrapping round the torus.

Our method does not reveal the excitation statistics.  However, the excitations of any non-degenerate gapped state at non-integer filling have fractional particle number by Laughlin's gauge argument \cite{fqh-frac-gauge}, which is still valid in the presence of the non-Abelian field, and hence fractional statistics \cite{fqh-charge-stat}.  This does not distinguish Abelian from non-Abelian statistics, but given the lack of a visible phase transition we expect the same statistics as the Laughlin state, i.e. Abelian $\theta=\pi/2$ (semions) \cite{laughlin-stat}.

\subsection{$\nu=1$: Moore-Read state}

\begin{figure*}
\includegraphics[width=15cm]{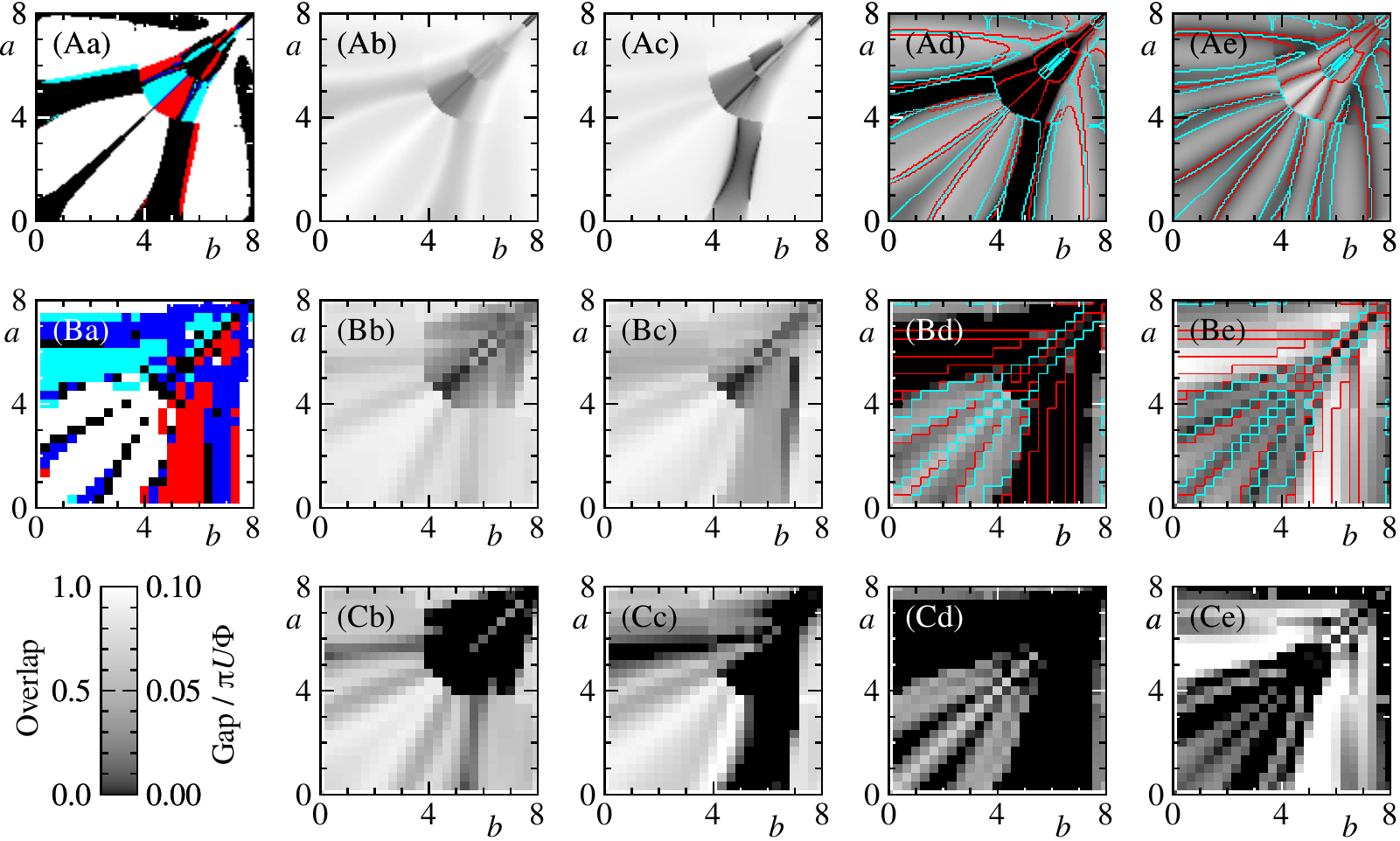}
\caption{$\nu=1$, (A) $N=8$ particles, (B) $N=12$ particles, (C) $1/N$ extrapolation to infinitely many particles, based on data for 6, 8, 10 and 12 particles.  Due to strong size dependency, this extrapolation is not reliable.  (a) Haldane momenta $\bi{K}$ below the largest energy gap in the lowest 10 states.  White: Moore-Read ($\bi{K}=(N/2,0),(0,N/2),(N/2,N/2)$).  Cyan (pale): Smectic-Y (two or more $\bi{K}$ with equal $K_y$ and evenly spaced $K_x$, suggesting stripe order parallel to $y$ \cite{4rr-numeric}).  Red (medium): Smectic-X (two or more $\bi{K}$ with equal $K_x$ and evenly spaced $K_y$).  Blue (dark): Single state (usually, but not always, $\bi{K}=(N/2,0)$ or $(0,N/2)$).  Black: any other combination.  (b,c) Overlap of the numerically calculated lowest $\bi{K}=(N/2,N/2)$ (b) or $\bi{K}=(0,N/2)$ (c) state with the projection of the exact Moore-Read state onto the lowest Landau level.  For the third component ($\bi{K}=(N/2,0)$), reflect (c) in the $a=b$ diagonal.  (d) Energy gap between the highest of the three Moore-Read states ($\bi{K}=(N/2,0),(0,N/2),(N/2,N/2)$) and the next lowest state (as shown in figure \ref{energies_k}c); negative (truncated to zero in the figure) if these three states are not the lowest three.  (e) Energy splitting between the highest and lowest of these 3.  Lines in (d,e) are energy level crossings of these 3 states, cyan if the ground state is involved, red if not.} \label{read_varyn}
\end{figure*}

The $p$th Read-Rezayi state for bosons is defined as the $\nu=p/2$ state which vanishes when any $p+1$ particles come together; it is unique on a disc but $p+1$-fold degenerate (including the twofold centre of mass degeneracy if $p$ is odd) on a torus \cite{rr-cluster}.  It is equal to the symmetrized product of $p$ Laughlin states \cite{rr-lprod}.  It is not an exact eigenstate for $p\neq 1$, but in Abelian field has been found to be a good trial state for small $p$ \cite{rr-torus-numeric}.

The $p=2$ ($\nu=1$) case is also called the Moore-Read or Pfaffian state.  Its excitations are Ising anyons, whose non-Abelian statistics have been tentatively detected in a related fermionic state \cite{mr-interference-expt}.  These are not computationally universal on their own \cite{mr-qc}, but can be made so with relatively noisy non-topological operations \cite{mr-qc,mr-qc2}.  On a torus the Moore-Read state is 3 states at Haldane momenta $\bi{K}=(N/2,0),(0,N/2),(N/2,N/2)$ \cite{rr-torus-numeric} (where $\bi{K}$ is scaled to take integer values $0,\ldots,N-1$); for a system to be in the Moore-Read phase, in the infinite size limit these 3 states must be topologically degenerate \cite{mr-torus-degen} and have an energy gap between them and all other states.  However, the Moore-Read state has a smaller energy gap and a longer correlation length than the Laughlin state, and hence is more sensitive to finite size error, which can split the degeneracy.  At the sizes we can access (up to 12 particles) it is not well converged, as shown by the strong system size dependence in figure \ref{read_varyn}.

We find a positive energy gap (i.e. the three lowest energy states are at the right three $\bi{K}$ for a Moore-Read state, figure \ref{read_varyn}d) and large overlaps between the exact ground state and the projected (into the lowest Landau level) Moore-Read state at all three $\bi{K}$ (figure \ref{read_varyn}b,c) over roughly the same region as the Laughlin state, i.e. $a,b\lesssim 4$ plus probably-artefact revivals when $a$ or $b$ is large.  However, the gap and degeneracy splitting (figure \ref{read_varyn}e) vary significantly within this region, with 2-3 ``good" stripes (large gap, small splitting, energy spectrum shown in figure \ref{energies_k}b), separated by ``bad" stripes (small gap, large splitting, figure \ref{energies_k}c).  Simple $1/N$ extrapolation suggests that only the ``good" stripes are in the Moore-Read phase (zero splitting in the infinite size limit, figure \ref{read_varyn}Ce), but the strong system size dependence makes this result unreliable, and it is a plausible alternative that the entire $a,b\lesssim 4$ region is in the Moore-Read phase.  The stripes arise from crossings of the three Moore-Read state energies (marked by lines in figure \ref{read_varyn}d,e); near triple crossings, giving a greatly reduced splitting, seem to be more common than would be expected by chance.

At stronger fields, the Moore-Read state is replaced by smectic (stripe ordered), non-degenerate gapped, or gapless states (figure \ref{read_varyn}a).  Which of these occur in what regions of $a,b$ parameter space is strongly system size dependent, and we hence do not know which, if any, of these survive in the infinite size limit.  As discussed above for general $\nu$, at these fields the width of a single particle eigenstate is a large fraction ($\gtrsim\frac{1}{2}$) of the the torus circumference, and strong finite size artefacts are hence likely.

We did not perform a similarly detailed study of the $\nu=3/2$ state (which if Read-Rezayi supports universal topological quantum computation \cite{rr-qc-braid}), as at the sizes we can reach, the two $\bi{K}=0$ states that should be its degenerate ground states are mostly not even the lowest two states (figure \ref{energies_k}d); this is consistent with previous Abelian-field results \cite{3rr-numeric}.  However, we note that it too can have its splitting reduced and its ``gap" made less negative by some choices of non-Abelian field, and that its components sometimes avoided cross (one such crossing causes the rapid decrease of overlap at $a\approx 2$ in figure \ref{density2}b), suggesting similar behaviour to $\nu=1$.
\subsection{State-dependent atomic interactions}
The above assumes equal inter- and intraspecies interaction, $U_{\uparrow\uparrow}=U_{\downarrow\downarrow}=U_{\uparrow\downarrow}$.  For hyperfine states of real atoms, these are usually nearly but not exactly equal, and they can be far from equal near a Feshbach resonance.  Also, an alternative method of generating the U(2) field \cite{u2f-valley} uses the two pseudo-internal states (momentum valleys) that occur at $\Phi\approx 1/2$, which have a fixed $1:1:2$ interaction ratio \cite{fqh-high-flux}, instead of true internal states.

We tested the effect of asymmetric interactions by repeating some of the above calculations for $U_{\uparrow\uparrow}:U_{\downarrow\downarrow}:U_{\uparrow\downarrow}=1.024:0.973:1$ (the $\ket{F,m_F}=\ket{1,-1},\ket{2,1}$ states of $^{87}$Rb \cite{rotating-gas-review}), $1:1:0$ (i.e. no interspecies interaction) and $1:1:2$.  The small deviation from 1 had no visible effect; since there is only one lowest Landau level, not a degenerate pair, there is nothing to phase separate.  The larger increase (decrease) of interspecies interaction increased (decreased) the parameter range over which FQH was seen and its energy gap, but did not alter its qualitative features.
\section{Conclusions}\label{conc}

We numerically studied the fractional quantum Hall effect in bosons in a U(2) gauge field.  For moderate non-Abelian field strengths, we find that it behaves similarly to a \emph{one} internal state Abelian-field quantum Hall system, despite actually having two internal states.  Within this regime, the energy gap of the non-Abelian anyon states $\nu=1,3/2$ is strongly dependent on the non-Abelian field parameters, with some settings giving larger gaps than in the true one internal state system.  However, as the gap is also strongly dependent on the system size, it is not clear whether this effect persists in the infinite size limit.

In strongly asymmetric non-Abelian fields, the system reverts to behaving as a \emph{two} internal state quantum Hall system (i.e. like it would without any non-Abelian field), while in strong near-symmetric non-Abelian fields, it does not exhibit the fractional quantum Hall effect at all.

\ack We thank Michele Burrello for useful discussions.  R.N.P. acknowledges financial support from the European Commission of the European Union under the FP7 STREP Project HIP (Hybrid Information Processing).
\section*{References}
\bibliographystyle{apsrev4-1}
\bibliography{references}
\end{document}